# Ultrafast Third-Order Nonlinear Optical Response of Charge Coupled Gold Nanoparticle-Ge$_{24}$Se$_{76}$ Heterostructure

Vinod Kumar[1] Rituraj Sharma[1], Abhishek Bhatt[1], I. Csarnovics[2], Petr Nemec[3], H. Jain[4], and K. V. Adarsh[1]*

[1]*Department of Physics, Indian Institute of Science Education and Research, Bhopal 462066, India*

[2]*Department of Experimental Physics, Institute of Physics, Faculty of Science and Technology, University of Debrecen, 4032 Debrecen, Hungary*

[3]*Department of Graphic Arts and Photophysics, Faculty of Chemical Technology, University of Pardubice, 53210 Pardubice, Czech Republic*

[4] *Department of Materials Science and Engineering, Lehigh University, Bethlehem, Pennsylvania 18015*

The donor-acceptor interaction of a charge-coupled heterostructure encompassing a metal and an amorphous semiconductor subjected to a laser field has many potential applications in the realm of nonlinear optics. In this work, we fabricate an electron donor gold nanoparticle (AuNP) and acceptor amorphous Ge$_{24}$Se$_{76}$ heterostructure on a quartz substrate using a sequential thermal evaporation technique. In this charge coupled heterostructure, we demonstrate the ultrafast third-order nonlinear absorptive and refractive response, and their sign reversal compared to pristine Ge$_{24}$Se$_{76}$. Enhanced optical nonlinearity in these heterostructures of varying plasmonic wavelengths is due to charge transfer, verified by the Raman spectroscopy. Further, the ultrafast transient absorption measurements support the thesis of charge transfer in the AuNP/Ge$_{24}$Se$_{76}$ heterostructure. These findings open up exciting opportunities for developing novel device technologies with far-reaching applications in nonlinear optics.

# Introduction

Metal-semiconductor heterostructures have attracted significant attention owing to their unique optoelectronic properties and potential applications [1–3]. These heterostructures offer tunable and versatile platforms for various optoelectronic devices, including ultrafast optical limiters [4,5] and passive mode-locking saturable absorbers for ultrashort laser pulses [6,7]. The interaction between donor and acceptor components in these heterostructures, typically involving metals and semiconductors, has been a subject of extensive investigation due to the creation of a space-charge region, enabling the design of multifunctional properties not achievable in individual materials [8-10]. Over the years, many metal-semiconductor heterostructures have been realized, such as metal-core/semiconductor-shell structures and non-epitaxially coupled metal-semiconductor hybrids (both organic and inorganic), demonstrating tailored optical properties for applications in optical limiters, photovoltaics, solar cells, and photodetectors [5,11-13]. The charge-transfer interaction in these heterostructures has proven to be a powerful mechanism for modifying their chemical and optoelectronic properties [14,15]. Notably, recent studies have reported exceptional enhancements in the third-order nonlinear optical response through efficient charge transfer from metal shells to semiconductor cores triggered by selective excitation at the plasmonic wavelength [16]. Similarly, charge-coupled systems like single-wall carbon nanotube-$VSe_2$ have exhibited remarkable enhancements in the nonlinear optical response across a broad spectral range (400 – 900 nm) [17]. Large enhancements in third-order nonlinear optical responses have also been observed in heterostructures such as $MoSe_2$/graphene oxide [8], gold nanoparticles (AuNP)/reduced graphene oxide [5], gold-cadmium sulfide [18], and gold-cadmium telluride [19] systems. Given these advancements, it becomes crucial to investigate the impact of charge transfer on the ultrafast third-order nonlinear optical response in the $AuNP/Ge_{24}Se_{76}$ heterostructure, particularly in the local field environment of the surface plasmon resonance of the metal nanoparticles. Understanding the intricate dynamics and optoelectronic behavior in this heterostructure will provide valuable insights for harnessing its potential in nonlinear optics and pave the way for future applications.

To address this challenge, we have used a thermal evaporation process to synthesize a charge-coupled $AuNP/Ge_{24}Se_{76}$ heterostructure with various plasmonic wavelengths. By optimizing the plasmonic wavelength to coincide with the optical bandgap of $Ge_{24}Se_{76}$, we investigated the charge transfer between AuNP and $Ge_{24}Se_{76}$. This article presents our findings on ultrafast third-order nonlinear absorptive and refractive response in the charge-coupled $AuNP/Ge_{24}Se_{76}$ heterostructure using Z-scan measurements in the strong coupling regime facilitated through charge transfer. Remarkably, we have observed a sign reversal in the ultrafast nonlinear optical response of the heterostructure compared to pristine $Ge_{24}Se_{76}$. While $Ge_{24}Se_{76}$ exhibits weak two-photon absorption (TPA) and self-focusing (positive nonlinearity), the strongly coupled heterostructure demonstrates saturable absorption (SA) and self-defocusing (negative nonlinearity). Moreover, we observe a significant enhancement in the magnitude of the saturation intensity and nonlinear refractive index of $AuNP/Ge_{24}Se_{76}$ at various surface plasmon resonance wavelengths. For instance, the heterostructures exhibit a large saturation intensity of $98\pm9$ $GW/cm^2$ and a four-fold enhancement in the nonlinear refractive index, which has an opposite sign compared to $Ge_{24}Se_{76}$.

## Experimental details

**Raman spectroscopy:** The Raman spectra were acquired using a Horiba LabRAM high-resolution spectrometer equipped with a 632.5 nm He-Ne laser as the excitation source. To prevent sample damage, the laser intensity at the sample surface was maintained at a low level ($< 1$ mW).

**Z-scan measurement:** For the Z-scan measurements, we employed a Ti: Sapphire Regenerative Amplifier System. The excitation source consisted of 150 fs pulses with an energy equivalent to 0.68 times the bandgap energy ($E_g$) of $Ge_{24}Se_{76}$. The femtosecond laser pulse was focused onto the sample using a plano-convex lens with a focal length of 30 cm. The sample position was controlled along the Z-axis using a motorized translation stage. In our experimental setup, the Rayleigh length ($Z_0$) was 1.8 mm, and the beam waist ($W_0$) was approximately 50 µm.

**Ultrafast transient absorption measurements:** In the ultrafast transient absorption measurements, an 800 nm laser beam with 120 fs pulses, 1 kHz repetition rate was employed. To generate the pump pulse, the initial laser beam was directed through an optical parametric amplifier (TOPAS), resulting in a 545 nm pump pulse with a fluence of 500 µJ/cm$^2$. The second pulse, serving as the probe pulse, was time-delayed using a computer-controlled translational stage and then focused onto a $CaF_2$ crystal plate generating white light spanning from 400-1000 nm. Further the acquired traces were chirp-corrected using a previously reported technique [20].

## Results and Discussion

We fabricated a strongly coupled AuNP/$Ge_{24}Se_{76}$ heterostructure using a thermal evaporation technique [21-23]. Initially, a 15 nm thick gold layer was deposited onto a quartz substrate, which was then transformed into Au nanoparticles (AuNP) via Ostwald ripening by heating the film at various temperatures ranging from 350 to 550 °C [22-24]. This process resulted in different nanoparticle sizes with corresponding plasmonic wavelengths. For example, nanoparticles of sizes 60, 80, and 100 nm exhibited plasmon wavelengths of 545, 580, and 620 nm, respectively (Fig. 1(a)). The size and morphology of the AuNP were characterized using an atomic force microscope (AFM), and more details can be found elsewhere [21]. Subsequently, a 450 nm thick $Ge_{24}Se_{76}$ film was thermally evaporated over the AuNP to form the strongly coupled AuNP/Ge$_{24}$Se$_{76}$ heterostructure. To investigate the interaction between the AuNP and the pristine Ge$_{24}$Se$_{76}$ domains within the heterostructures, we performed conventional ultraviolet-visible dual-beam spectroscopy and measured the absorption spectra of the samples.

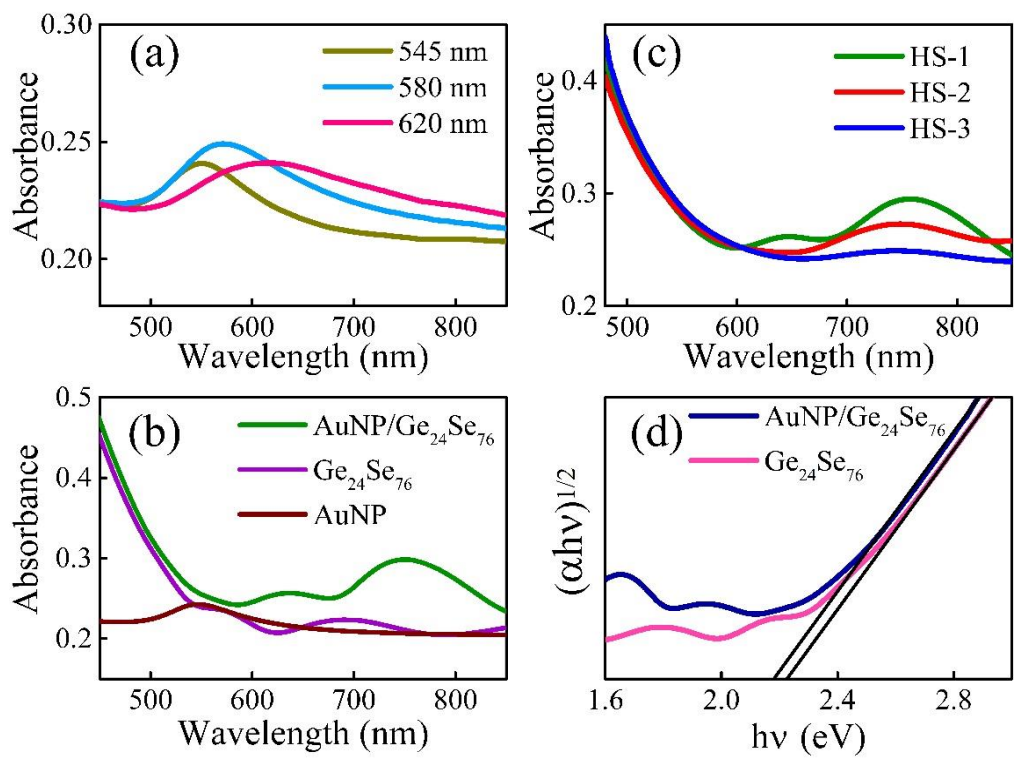

**Fig. 1.** (a) Optical absorption spectrum of AuNP revealing distinct plasmon bands at different wavelengths depending on the nanoparticle size. Specifically, nanoparticles with sizes of 60, 80, and 100 nm exhibit plasmon wavelengths of 545, 580, and 620 nm, respectively. (b) Optical absorption spectrum of AuNP/Ge$_{24}$Se$_{76}$ (HS-2) heterostructure, Ge$_{24}$Se$_{76}$ and AuNP. (c) Optical absorption spectrum of AuNP/Ge$_{24}$Se$_{76}$ heterostructures at various plasmon wavelengths. It can be observed that the overall spectrum of the heterostructure is redshifted compared to pristine Ge$_{24}$Se$_{76}$, indicating charge transfer and strong coupling between the AuNP and Ge$_{24}$Se$_{76}$ (d) The Tauc plot of AuNP/Ge$_{24}$Se$_{76}$ heterostructure (HS-2) and pristine Ge$_{24}$Se$_{76}$.

Figure 1(a) displays the optical absorption spectrum of AuNP with a size of 60 nm, revealing a plasmon band centered at a wavelength of 545 nm. Notably, as the particle size increases to 80 and 100 nm, the plasmon wavelength shifts to longer wavelengths of 580 and 620 nm, respectively. Figure 1(b) shows the optical absorption spectra of AuNP/Ge$_{24}$Se$_{76}$ heterostructure and pristine $Ge_{24}Se_{76}$. Both spectra exhibit a characteristic interference pattern, with the $Ge_{24}Se_{76}$ layer demonstrating lower absorbance levels compared to the $AuNP/Ge_{24}Se_{76}$ heterostructure, indicating the effect of metallic nanoparticles on the optical properties. Furthermore, the optical absorption spectrum of the heterostructure suggests a superposition of signals from the individual components; however, the overall signal has exhibited a redshift equated to pure $Ge_{24}Se_{76}$, signifying charge transfer and strong coupling between the AuNP and $Ge_{24}Se_{76}$. Additionally, Figure 1(c) presents the optical absorption spectra of the heterostructure corresponding to three different plasmon wavelengths. It can be observed that the plasmon resonance undergoes a shift consistent with the size-dependent trend of metallic permittivity and plasmon oscillations [25,26]. Henceforth, we will refer to the $AuNP/Ge_{24}Se_{76}$ heterostructures as HS-1 (plasmon wavelength 545 nm), HS-2 (580 nm), and HS-3 (620 nm). We performed a Tauc plot analysis on the optical absorption spectra of the $AuNP/Ge_{24}Se_{76}$ heterostructures (HS-2) and $Ge_{24}Se_{76}$, as shown in Fig. 1(d) to find the optical bandgap. Our calculations revealed optical bandgap of $2.18 \pm 0.03$ and $2.24 \pm 0.02$ eV for $AuNP/Ge_{24}Se_{76}$ (HS-2) heterostructure and $Ge_{24}Se_{76}$, respectively. Interestingly, we observed that this bandgap value is in proximity to the surface plasmon resonance band of the AuNP. Strikingly, we observed 60 meV redshift in HS-2 compared to pristine $Ge_{24}Se_{76}$. This finding suggests a potential correlation between the optical bandgap of $Ge_{24}Se_{76}$ and the surface plasmon resonance wavelength of AuNP, indicating an opportunity for optimal coupling between the two materials.

To gain further insights into the coupling between AuNP and $Ge_{24}Se_{76}$, we conducted Raman spectroscopy measurements on both pristine samples and heterostructures. The resulting Raman spectra are presented in Fig. 2(a). The Raman spectrum of $Ge_{24}Se_{76}$ exhibits three distinct vibrational modes, each associated with specific bonding configurations. These modes include a 194 $cm^{-1}$ band attributed to corner-sharing vibrations of Ge-Se tetrahedra, a sideband around 210 $cm^{-1}$ associated to edge-sharing tetrahedra, and a broad peak at 256 $cm^{-1}$ originating from Se-Se bonds within the polymeric chain of Se atom [27-29]. The interaction between the constituents in the heterostructure leads to modifications in these band positions, resulting in both stiffening and softening effects [30].

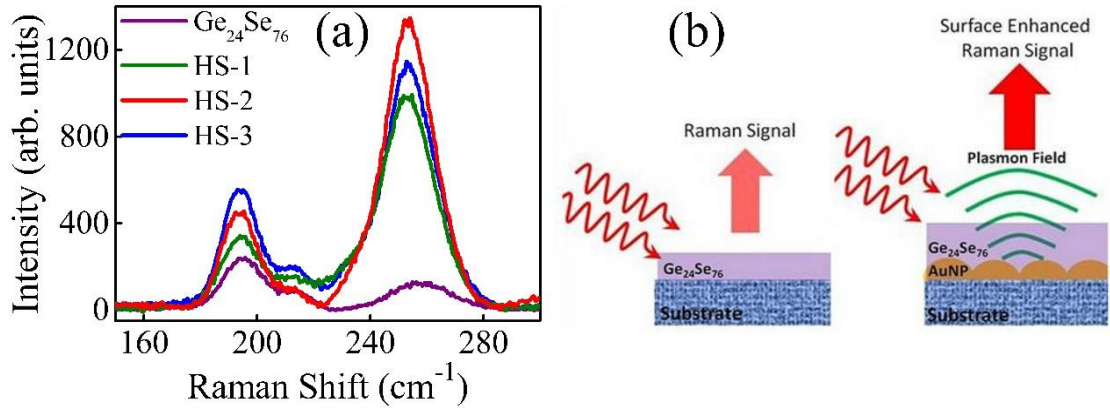

**Fig. 2.** (a) Raman spectrum of $Ge_{24}Se_{76}$ and AuNP/$Ge_{24}Se_{76}$ heterostructures. It is evident from the figure that the Raman spectrum of $Ge_{24}Se_{76}$ is significantly enhanced in the heterostructure. (b) Schematic showing the mechanism of plasmonic enhancement of Raman signals.

Notably, the intensities of the Ge-Se and Se-Se Raman contributions experience a significant enhancement in the heterostructure, reaching magnitudes up to five times higher than those in the pristine $Ge_{24}Se_{76}$. This enhancement can be ascribed to surface-enhanced Raman scattering facilitated by the presence of AuNP [31]. The interaction of light with the surface plasmon oscillations near the AuNP surface enables surface-enhanced Raman scattering to occur [32,33]. In this context, it is worth considering that the strong coupling and charge transfer between the AuNP and $Ge_{24}Se_{76}$ contribute to the observed enhancement. It is likely that only the symmetric vibrational modes of the sample are selectively enhanced through the Franck–Condon contribution, while both symmetric and nonsymmetric vibrations can be enhanced via the Herzberg–Teller effect [34]. Furthermore, we observe a similar enhancement of Raman signals across all heterostructures. Figure 2(b) presents a schematic illustrating the process of plasmonic enhancement of the Raman signal. When the frequency of the exciting radiation resonates with the plasmon oscillation frequency, there is a significant increase in the dipolar electric field at the surface of the metal. This enhanced electric field substantially boosts the Raman signal of $Ge_{24}Se_{76}$ located on the surface of the AuNP [31,32].

To investigate the ultrafast third-order nonlinear optical response of the AuNP/$Ge_{24}Se_{76}$ heterostructures and pristine $Ge_{24}Se_{76}$, we conducted both open and closed aperture Z-scan measurements. These measurements enabled us to examine the overall transmittance as a function of the sample position at a particular intensity [8,17]. In Figure 3(a), we present the open-aperture Z-scan results obtained at an on-axis peak intensity of 150 GW/$cm^2$ for the AuNP/$Ge_{24}Se_{76}$ heterostructures and pristine $Ge_{24}Se_{76}$, utilizing an 800 nm wavelength. In the case of pristine $Ge_{24}Se_{76}$, the Z-scan peak-shape trace demonstrates a reduction in transmission as it approaches the focal point, which is consistent with the two-photon absorption (TPA). It is important to note that this behavior is anticipated in $Ge_{24}Se_{76}$ due to the excitation at 800 nm falling within the two-photon excitation wavelength range. In stark contrast, the heterostructures exhibits saturable absorption (SA).

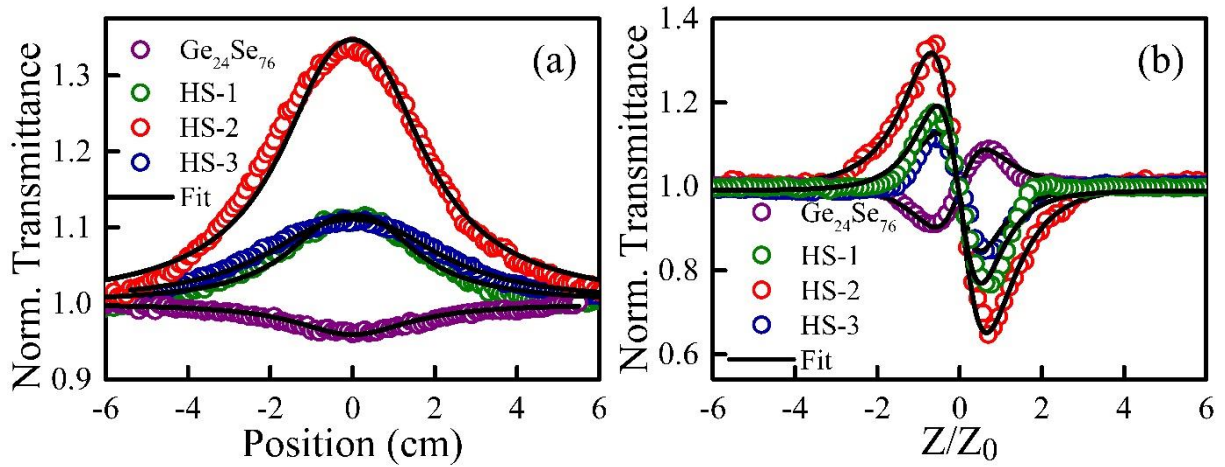

**Fig. 3.** Third-order nonlinear optical response of heterostructures and pristine $Ge_{24}Se_{76}$ at an on-axis peak intensity of 150 GW/$cm^2$ and 800 nm excitation. (a) Open aperture Z-scan traces (b) Closed aperture Z-scan traces. Solid lines represent the numerical fits.

To quantify the observed nonlinear saturation intensity ($I_s$) and TPA coefficient ($\beta$), we employed the following differential equation [8,17] describing the nonlinear absorption,

$$\frac{dI}{dz} = -\alpha(I)I \qquad (1)$$

where I represent the intensity and $\alpha(I)$ denotes the intensity-dependent absorption coefficient expressed as

$$\alpha(I) = \frac{\alpha_0}{I + \frac{I}{I_s}} + \beta I \qquad (2)$$

where $I_s$ and $\alpha_0$ are, the saturation intensity and linear absorption coefficient, respectively. For a pulsed Gaussian beam, we can solve Eq. (1), and we can write the normalized transmittance T in the Z-scan experiment [8] as

$$T = \left(\frac{1}{P_0\sqrt{\pi}}\right) \int_{-\infty}^{+\infty} \ln\left(1 + P_0 e^{-t^2}\right) dt \qquad (3)$$

Where $P_0$ is given by $P_0 = \left[\beta(1-R)I_0 L_{eff}\right]/[1 + (Z/Z_0)^2]$ R, $Z_0$, $I_0$, L and $L_{eff}$ are the surface reflectively, Rayleigh length, peak intensity, length and effective length of the sample ( $L_{eff} = (1 - e^{\alpha_0 L})/\alpha_0$). The parameters obtained from the optimal fit to the normalized transmittance data are presented in Table I. We can observe from Table I that the β value for pristine $Ge_{24}Se_{76}$ is determined to be β = (60 ± 5) ×$10^{-2}$ cm/GW. Notably, the Z-scan trace of the AuNP did not exhibit any nonlinear behavior under the 800 nm excitation. However, the Z-scan peak-shape responses of the heterostructures displayed a remarkable increase in transmittance, indicating SA response. This behavior starkly contrasts the TPA observed in $Ge_{24}Se_{76}$ and the absence of nonlinearity in the AuNP. The saturation intensity ($I_s$) for the heterostructure (HS-2) is determined to be $I_s$ = (98 ± 9) GW/cm$^2$ at the surface plasmon resonance wavelength of 580 nm, which is ∼ 50 times larger than pristine $Ge_{24}Se_{76}$ in the femtosecond regime. This finding demonstrates the significant transition from TPA to saturable absorption (SA) in the heterostructure. Additionally, the calculated ground state absorption cross-section (σ$_{GS}$) and excited state cross-section (σ$_{ES}$) (shown in Table I) support the experimental evidence of the SA response of the samples [17]. According to the condition for observing SA, the $\sigma_{GS}$ must be greater than the $\sigma_{ES}$, [8,17] and this condition is satisfied in all heterostructures. However, in the case of pristine Ge$_{24}$Se$_{76}$ $\sigma_{GS}$ is less than the $\sigma_{ES}$ resulting in TPA. We assume the larger value of $\sigma_{GS}$ in heterostructure is due to charge transfer. Interestingly, Table I reveals no clear trend in the saturation intensity values concerning the plasmon wavelength.

**Table I:** Nonlinear optical parameters: TPA Coefficient ($\beta$), Saturation Intensity ($I_s$) and nonlinear refractive index (n$_2$), $\sigma_{GS}$ and $\sigma_{ES}$.

| Sample | $\beta$ (10$^{-2}$ cm/GW) | $I_s$ (GW/cm$^2$) | $\sigma_{GS} \times 10^{-20}$ (cm$^2$) | $\sigma_{ES} \times 10^{-20}$ (cm$^2$) | n$_2$ (10$^{-6}$ cm$^2$/GW) |
|---|---|---|---|---|---|
| Ge$_{24}$Se$_{76}$ | 60 ± 5 | 2.0 ± 0.4 | 1.5 ± 0.3 | 3.5 ± 0.4 | 1.5 ± 0.2 |
| Au | - | - | - | - | - |
| HS-1 | -(5.0 ± 0.4) | 48 ± 4 | 4.8 ± 0.2 | 1.6 ± 0.2 | -(2.7 ± 0.3) |
| HS-2 | -(2.0 ± 0.3) | 98 ± 9 | 8.6 ± 0.8 | 2.5 ± 0.5 | -(6.0 ± 0.5) |
| HS-3 | -(4.0 ± 0.5) | 42 ± 3 | 4.5 ± 0.5 | 1.5 ± 0.2 | -(2.5 ± 0.4) |

After illustrating the nonlinear absorption using the open aperture Z-scan, we conducted a closed aperture Z-scan measurement to demonstrate the magnitude and sign of the nonlinear refractive index of the heterostructure and pristine Ge$_{24}$Se$_{76}$ in the femtosecond regime. We estimated the nonlinear refractive index (n$_2$) value by fitting the closed aperture Z-scan traces with the theoretical equation [35],

$$T_N(y) = 1 + \frac{1}{\sqrt{2}} \frac{4\,y\,\varphi_0 - (y^2+3)p_0}{(y^2+1)(y^2+9)}$$

$$+ \frac{1}{\sqrt{3}} \frac{4\,{\varphi_0}^2(3y^2-5) + {p_0}^2(y^4+17y^2+40) - 8\varphi_0 p_0 y(y^2+9)}{(y^2+1)^2(y^2+9)(y^2+25)} \qquad (4)$$

where $T_N$ is the normalized transmittance, y = Z/Z$_0$ is the sample position, $p_0 = \beta I_0(1-R)L_{eff}$ is the phase shift, $\varphi_0 = 2\pi n_2 I_0(1-R)L_{eff}/\lambda$ is the on-axis phase shift with n$_2$. Figure 3(b) showcases the closed aperture Z-scan traces for the heterostructures and $Ge_{24}Se_{76}$ at a moderate peak intensity of 150 GW/cm² and excitation wavelength of 800 nm. At this modest intensity, we observed a valley in the pre-focal position followed by a peak in the post-focal position, i.e. (valley-peak) indicating a positive refractive nonlinearity in pristine $Ge_{24}Se_{76}$. This behavior suggests a self-focusing nature of the material with an $n_2$ value of $(1.5 \pm 0.2) \times 10^{-6}$ cm²/GW. This positive refractive nonlinearity can be attributed to bound carriers generated due to nonresonant excitation. However, a surprising observation was made for the $AuNP/Ge_{24}Se_{76}$ heterostructures, where the Z-scan signature switched to a peak-valley structure. This indicates a negative refractive nonlinearity, leading to the self-defocusing nature of the material. The HS-2 exhibited an $n_2$ value of $-(6.0 \pm 0.5) \times 10^{-6}$ which is $\sim 50$ times of $Ge_{24}Se_{76}$. The measured $n_2$ values are summarized in Table I, which indicates that the $AuNP/Ge_{24}Se_{76}$ heterostructure shows a higher $n_2$ compared to the pristine $Ge_{24}Se_{76}$. Notably, the negative value of $n_2$ can be ascribed to the predominance of free carriers in the system [35]. This analysis allows us to quantitatively determine the nonlinear refractive properties of the samples and provides valuable insights into the interaction between the AuNP and $Ge_{24}Se_{76}$ in the heterostructure.

To gain a comprehensive understanding of the experimental findings pertaining to the charge transfer observed through third-order nonlinear optical measurements in the charge-coupled $AuNP/Ge_{24}Se_{76}$ heterostructures, we carried out femtosecond transient absorption (TA) measurements. To illustrate these results, we present Figure 4(a), which showcases the TA spectrum of AuNP at various probe delays. The spectrum indicates that the surface plasmon resonance peak aligns with the wavelength range demonstrating significant photobleaching. The observed bleach in the TA can be attributed to the broadening of the surface plasmon resonance caused by quadruple (non-dipolar) plasma oscillations induced by the pump pulse [36]. Figure 4(b) illustrates the TA spectrum of $Ge_{24}Se_{76}$, encompassing the spectral range from the bandgap to the sub-bandgap regions.

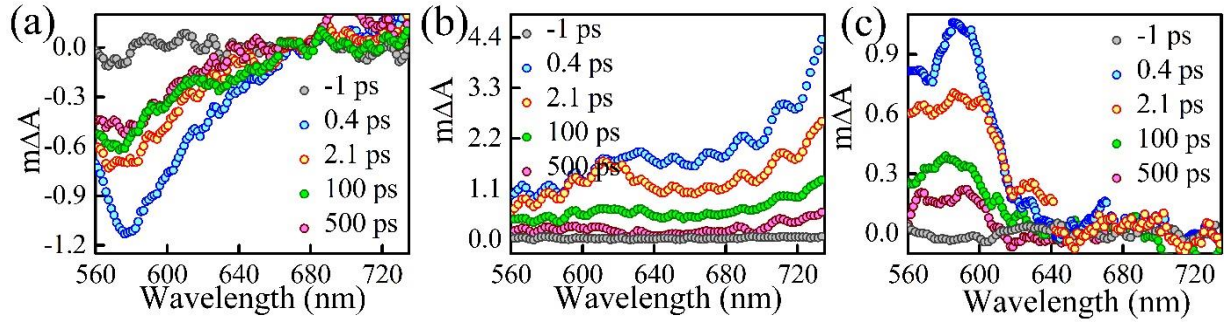

**Fig. 4.** Transient absorption spectrum of (a) AuNP (b) pristine $Ge_{24}Se_{76}$ and (c) HS-2 heterostructure at various probe delays.

The spectrum exhibits absorption maxima in the sub-bandgap region [37], providing compelling evidence for the self-trapped excitonic band [38,39]. Remarkably, the TA spectrum of HS-2 (Fig. 4(c)) exhibits a dramatic reduction in the TA signal within the sub-bandgap region. This observation provides compelling evidence for the strong coupling between self-trapped excitons and localized surface plasmons in the AuNP/$Ge_{24}Se_{76}$ heterostructures.

In order to deliver a comprehensive explanation for the observed quenching of TA in the heterostructures, we consider that the photoexcited electrons are effectively transferred from AuNP to $Ge_{24}Se_{76}$, which hinders the self-trapping of excitons and results in substantial reduction of photoinduced absorption in the below bandgap region of the $Ge_{24}Se_{76}$. This hypothesis is reinforced by empirical evidence, which demonstrates that the plasmon absorption signal (i.e., the AuNP photobleaching signal) is also quenched. Specifically, upon irradiation by the pump beam, the photoexcited electrons transfer from AuNP to $Ge_{24}Se_{76}$, resulting in the accumulation of a negative charge on the latter. Figure 5(a-c) displays contour plots of TA spectra for the three heterostructures, HS-1, HS-2, and HS-3. Strikingly, it is evident from the figure that the magnitude of photoinduced absorption reduces significantly as the plasmonic wavelength reaches near the bandgap of pristine $Ge_{24}Se_{76}$.

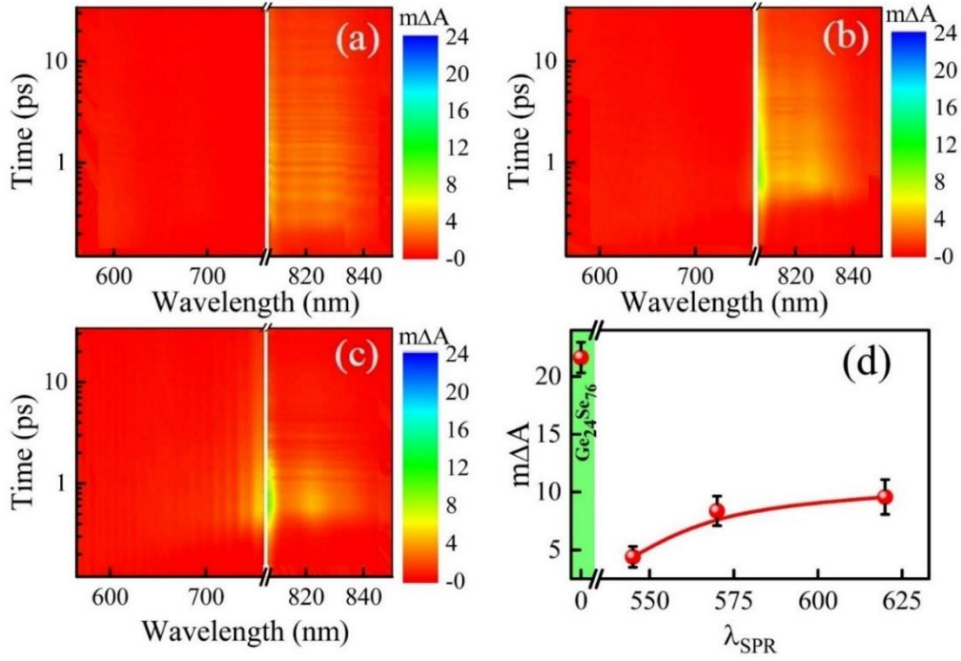

**Fig. 5.** Contour plots of TA spectra for (a) HS-1 (b) HS-2 and (c) HS-3 (d) Comparison of change in absorption corresponding to excitonic maxima ($\lambda = 807$ nm) for different HS. Surface plasmon resonance ($\lambda_{SPR}$) =0 corresponds to pristine $Ge_{24}Se_{76}$.

Furthermore, Fig. 5(d) presents a plot of the photoinduced absorption maxima at 807 nm as a function of the plasmon wavelength. This plot highlights that the most effective exciton quenching occurs when the surface plasmon resonance reaches the bandgap. Thus, the TA quenching in $AuNP/Ge_{24}Se_{76}$ heterostructures can be precisely controlled by regulating the plasmonic wavelength, which is determined by the size of the AuNP. The plasmon wavelength dependence of transient absorption suggests that the size of the AuNP directly impacts the charge transfer.

## Conclusions

In summary, we demonstrate the third-order nonlinear optical response in charge coupled $AuNP/Ge_{24}Se_{76}$ heterostructures within the strong coupling regime enabled by the charge transfer. We have observed a sign reversal of ultrafast third-order nonlinear absorption and refraction in heterostructures compared to pristine $Ge_{24}Se_{76}$. For instance, in sharp contrast to the weak two-photon absorption (TPA) and self-focusing (positive nonlinearity) of $Ge_{24}Se_{76}$ the strongly coupled heterostructure shows saturable absorption (SA) and self-defocusing (negative nonlinearity). Moreover, the enhanced blue shifted Raman signal and narrow spectral width of the ultrafast transient absorption and quenching of localized surface plasmon absorption in the heterostructure supports the thesis of the charge transfer between pristine $Ge_{24}Se_{76}$ and AuNP. All these studies clearly show that the coupling between amorphous materials and metal nanoparticles can generate a wide range of exciting opportunities for novel optical functionalities not exhibited by the individual components. These developments hold tremendous potential for developing new device technologies, such as all-optical data processing, data storage, passive mode-lock saturable absorber and semiconductor devices.

## Acknowledgements

The authors thank the Science and Engineering Research Board (CRG/2019/002808); the Department of Science and Technology, Ministry of Science and Technology, India (DST-FIST project (PSI-195/2014) for financial support. Istvan Csarnovics acknowledges the János Bolyai Research Scholarship from the Hungarian Academy of Sciences (BO/348/20). The New National Excellence Program's ÚNKP-22-3-II-DE-53 and ÚNKP-22-5-DE-407 also provided support from the National Research, Development, and Innovation Fund. The US National Science Foundation is acknowledged for supporting (DMR-2123131) this international collaboration that was established through International Materials Institute for New Functionality in Glass (DMR-0844014)